# A novel imaging setup for hybrid radiotherapy tailored PET/MR in patients with head and neck cancer

**Short running title:** A novel radiotherapy setup for PET/MR

René M Winter, PhD[1] • Ola Engelsen, PhD[2] • Ole Jakob Bratting, MSc[1] • Njål Brekke, PhD[3] • Jostein Sæterstøl, MSc[3] • Oddbjørn Sæther, PhD[4] • Kathrine R Redalen, PhD[1]

[1] Department of Physics, Norwegian University of Science and Technology, Trondheim, Norway

[2] PET Imaging Center, University Hospital of North Norway (UNN), Tromsø, Norway

[3] Center for Nuclear Medicine/PET, Haukeland University Hospital, Bergen, Norway

[4] Department of Radiology and Nuclear Medicine, St. Olavs hospital, Trondheim University Hospital, Trondheim, Norway

**Corresponding author:** René Winter, rene.winter@ntnu.no

**Statistical analysis:** René Winter, rene.winter@ntnu.no

**Conflict of interest:** This work was supported by the Norwegian Cancer Society under grant no. 198116-2018, and by Trond Mohn Foundation under grant no. TMS2019TMT08 to the Norwegian Nuclear Medicine Consortium. Grants funded postdoc position and data acquisition. RW received travel grants via the Norwegian University of Science and Technology from the Norwegian Nuclear Medicine Consortium, which is funded by Trond Mohn Foundation. OE received support from The Norwegian Pedriatic Cancer Society ("Barnekreftforeningen"). The PET-MR and associated radiotherapy equipment at the University Hospital North Norway in Tromso are gifts from philantropist Trond Mohn, Bergen, Norway.

**Funding:** This work was supported by the Norwegian Cancer Society under grant no. 198116-2018, and by Trond Mohn Foundation under grant no. TMS2019TMT08 to the Norwegian Nuclear Medicine Consortium.

**Data availability:** The data that support the findings of this study are stored in an institutional repository and are available from the corresponding author upon reasonable request. This excludes in-vivo data containing information that could compromise the privacy of research participants.

**Acknowledgements:** The authors thank Kirsten Margrete Selnæs from St Olavs hospital, Trondheim, for her valuable support with ACR phantom experiments.

# A novel imaging setup for hybrid radiotherapy tailored PET/MR in patients with head and neck cancer

## Abstract

**Purpose:** Radiotherapy commonly relies on CT, but there is growing interest in using hybrid PET/MR. Therefore, dedicated hardware setups have been proposed for PET/MR systems which enable imaging in radiotherapy treatment position. These radiotherapy setups typically include a flat tabletop, positioning tools and coil holders specifically tailored to the devices. However, reduced MR image quality has been reported. Especially in neck and upper thorax, conventional radiotherapy setups are not optimal as they consist of head-only coil configurations. The purpose was to develop a novel PET/MR radiotherapy setup for improved MR image quality in head, neck and thorax and to test compliance in a multicenter setting.

**Methods:** A novel radiotherapy setup was designed, prototyped and tested on a 3T PET/MR system in three different centers. Imaging experiments were conducted in phantoms and healthy volunteers to compare against a standard radiotherapy setup. Imaging protocols included T1-, T2-, and diffusion-weighted MR (DWI). Finally, compliance with American College of Radiology (ACR) and the Quantitative Imaging Biomarker Alliance (QIBA) acceptance criteria was evaluated.

**Results:** SNR in neck/thorax was increased by a factor of 1.6 in phantom (p = 0.031) and volunteer images alike. The new setup passed ACR detectability and QIBA SNR tests, which the standard setup failed. The new setup passed all but two ACR test criteria in the three centers, presented repeatability and reproducibility variations of 4.9% and 7.8% and met all QIBA criteria for DWI except ADC precision.

**Conclusion**: The proposed setup yielded significantly higher SNR, better detectability, and complied with nearly all ACR and QIBA image quality criteria. It may thus advance the usage of PET/MR for radiotherapy purposes.

# 1. Introduction

Image-guidance in radiotherapy is conventionally based on X-ray computed tomography (CT), but the interest in integrating positron emission tomography (PET) and magnetic resonance (MR) imaging is steadily growing (1–3). This is especially the case for head and neck cancer, where the response to radiotherapy is still resulting in high rates of loco-regional recurrence, and early and late side-effects (4). PET and MR are attractive modalities to guide and ultimately improve radiotherapy by providing high soft tissue contrast and structural information combined with multidimensional functional information, for example on tumor glucose metabolism (5), hypoxia (6) or diffusion (7). Studies have shown that such data can be highly valuable for a wide range of radiotherapy tasks such as target volume delineation (8,9), treatment response assessment (10), outcome prediction (11), patient stratification (12) and the development of imaging biomarkers for treatment adaptation (13,14).

The many potential applications make hybrid PET/MR, where PET and MR are combined in a single system and data can be acquired simultaneously, a highly promising modality for radiotherapy (15–18). In particular, recent advances in machine learning and artificial intelligence have led to the possibility of MR-only based radiotherapy workflows where an MR-derived synthetic CT replaces the traditional CT for radiotherapy planning and dose calculation (19,20). This elicits an enormous potential to advance the role of hybrid PET/MR to a one-stop shop for radiotherapy planning, as Olin et al. (21) and Ahangari et al. (22) recently demonstrated in head/neck and cervical cancer.

However, imaging for radiotherapy purposes requires positioning the patient for the imaging examination in the same way as for the radiation treatment (23). This involves the use of dedicated positioning hardware such as a flat tabletop and, for patients with head and neck cancer, head, neck and shoulder mask fixation devices and positioning aids. A special coil setup should snugly fit the radiotherapy equipment, so typically flexible body matrix coils are used. Achieving good MR image quality with such radiotherapy-tailored imaging setup on a hybrid PET/MR system is challenging. The coils should not be positioned directly on the patient in order to avoid deformation of the body anatomy for dose planning. In addition, since both the radiotherapy hardware and the MR coil setup attenuate the PET signal, predefined attenuation maps of the hardware should be stored on the system for the specific hardware location, and be used for PET signal attenuation correction (24). This demands a reproducible setup design with coil holders or coil bridges where the coils are more distant to the patient than during diagnostic scans. This impacts MR image quality (18,25,26). Different setup designs have been proposed for different PET/MR systems (21,27–31) and reductions in MR signal-to-noise ratio (SNR) in the range of 25% to 46% have been reported (27–30). Reduced MR image quality in terms of low-contrast object detectability scores have been documented (30). Two radiotherapy setups have been proposed specifically for imaging patients with head and neck cancer (21,27,28,31). However, the setup designs leave room for improvement for acquiring high-quality MR images in the neck and upper thorax region. This is particularly relevant for treatment planning of laryngopharyngeal cancer or lymph node metastases. The currently available setups all consist of a head-only coil setup used for both head, neck and thorax. An additional flex coil can be added for neck/thorax by placing it directly on the patient, but then the coil will not be correctable in PET signal attenuation correction (21). Other radiotherapy setups have been proposed for pelvic or whole-body imaging which might in principle be suitable to image head/neck and thorax (29,32), but feasibility and compatibility with neck and shoulder mask fixation has not been studied.

Therefore, we developed a novel radiotherapy prototype setup for clinical PET/MR imaging facilitating better MR SNR and image quality in neck and thorax, without making compromises in radiotherapy mask fixation or reproducible hardware positioning. The purpose was to advance PET/MR-based treatment planning and quantitative imaging biomarkers. The primary aim was to systematically investigate the MR image quality with this novel prototype setup in comparison to a typical head-only radiotherapy setup.

Multicenter trials are essential in the development of imaging biomarkers for personalized treatment of cancer but the use of different imaging systems may add variation to the imaging performance (33). To our knowledge, multicenter trials investigating MR image quality in radiotherapy imaging setups on hybrid PET/MR systems do not exist. Therefore, it was a second aim to test compliance of MR image quality with the prototype setup across different institutions.

## 2. Material and methods

### Study design

A novel radiotherapy setup was designed and prototyped in collaboration with Medibord Limited (Nottingham, UK), and tested in three different centers in Norway: University Hospital of North Norway in Tromsø (center 1), St. Olavs Hospital in Trondheim (center 2), and Haukeland University Hospital in Bergen (center 3). All three centers are part of a multicenter imaging trial in head and neck cancer (EMINENCE trial, clinicaltrials.gov no.

NCT04612075). Scans were performed on a 3T PET/MR system (Biograph mMR, Siemens Healthineers, Erlangen, Germany). Each PET/MR system was equipped with two radiotherapy setups: the novel prototype setup featuring an extended head/neck/thorax coil configuration, and a standard setup featuring a head-only coil configuration (Medibord).

To analyze the MR image quality with the new prototype setup, the study was divided into four parts. First, measurements were performed in one center with both prototype and standard setup using the large American College of Radiology (ACR) phantom. The aim was to investigate whether the prototype setup offers better SNR and image quality in neck and thorax positions compared to the standard setup, and whether it fulfills the acceptance criteria of the ACR society. Second, monthly ACR phantom tests were performed with the prototype setup in one center over three months to determine repeatability. Replicate measurements were made in the other two centers to investigate the intercenter reproducibility. Third, test measurements with the prototype and standard setups were performed in one center using the Quantitative Imaging Biomarker Alliance (QIBA) diffusion phantom (QalibreMD Diffusion Standard Model 128, High Precision Devices, Inc., Boulder, CO, USA). The aim was to investigate the quality of DWI in a neck position, and to test compliance with the quality acceptance criteria of the QIBA society. Last, clinical feasibility of the prototype setup was tested in healthy volunteers.

**Experimental setups**

The standard head-only radiotherapy setup consisted of a flat tabletop, a set of two head coil holders with two 6-channel mMR body matrix coils, and a baseplate for thermoplastic mask fixation (Figure 1a-d). The tabletop placement allows for using the system's spine coil array for posterior neck/thorax coverage. The new prototype setup consists of the same tabletop, head coil and baseplate setup, but in addition, a novel arc-shaped coil bridge was developed for using a third body matrix coil for the anterior neck and thorax (Figure 1e-f). The neck/thorax coil bridge was designed with low photon attenuation and to be compatible with the baseplate and head, neck and shoulder mask fixation devices, while providing reproducible coil and holder positioning for PET signal attenuation correction. The spine coil array can still be used for posterior neck/thorax signal coverage.

**ACR image quality tests**

To investigate the potential benefit of the novel prototype setup compared to the standard setup, MR test measurements were performed with both setups in one center (center 1) using the ACR phantom. The experimental setups are depicted in Supplementary Figure 1a-b. The ACR phantom is a hollow cylinder of acrylic plastic (length, 148 mm; diameter, 190 mm) filled with a nickel chloride and sodium chloride solution. It contains different structures designed to facilitate a variety of scanner performance tests. For the first set of MR test scans, a sagittal localizer and T1-weighted spin-echo (T1w) sequence were used and sequence parameters were set according to the ACR guidelines (acraccreditation.org). The tests comprised scans at 15 equidistant positions in superior-inferior (SI) direction to investigate the performance profile from head to thorax. The distance between positions was 4 cm. The mid position (0 cm) was set to the location of the patient's chin, which is at the intersection of head and neck/thorax coil parts (Supplementary Figure 1c-d). Phantom positioning was controlled using the laser system of the PET/MR system. At the first five scan positions (brain to head), data were only acquired once with the head coil parts, as these are identical in both setups. At the last three thorax positions, scans with the standard setup were omitted, as the phantom would be placed too far outside the field-of-view of the head coils.

MR image quality metrics were analyzed according to ACR guidelines: High-contrast spatial resolution in anterior-posterior (AP) and left-right (LR) direction, slice thickness accuracy, slice position accuracy, percent image intensity uniformity (PIU), low-contrast object detectability, percent-signal ghosting, and geometric accuracy. Geometric accuracy was evaluated based on the phantom dimensions (the phantom's length in SI in the localizer, and diameter in AP, LR and diagonal directions in the T1w image). ACR criteria served as reference values. In addition, signal, noise and SNR were analyzed based on two repeat scans of the T1w image. An average signal image and a difference image (noise image) were calculated. Signal was defined as a region of interest (ROI)-based spatial mean pixel value in the average signal image, $\bar{S}_{avg-ima}$; noise as the ROI-based spatial standard deviation in the difference image, $\sigma_{diff-ima}$, divided by $\sqrt{2}$. SNR was then calculated as:

$$SNR = \sqrt{2} \; \frac{\bar{S}_{avg-ima}}{\sigma_{diff-ima}}. \tag{1}$$

95% confidence intervals for signal and $SNR$ were estimated using the ROI based spatial standard deviation for signal, and error propagation for SNR. A Wilcoxon signed-rank test was used to evaluate whether differences in signal and SNR between prototype and standard setup were significant at the 5% level.

**Repeatability and reproducibility**

ACR phantom measurements with the prototype setup were performed in one center (center 2) once a month for a three-month period to determine the repeatability of MR image quality. A localizer, T1w, and T2w sequence were used. Sequence parameters were set according to the ACR guidelines. The phantom was positioned in center position (0 cm) simulating a typical head/neck bed position. The same image quality metrics as listed above were evaluated according to the ACR guidelines. Repeatability was quantified by the within-object standard deviation over the monthly repeats, $\sigma_{3m}$, and interpreted relative to the ACR tolerance using a within-center coefficient of variation (wCV):

$$wCV = 100\% \frac{\sigma_{3m}}{c_{ACR}}, \tag{2}$$

where $c_{ACR}$ is the ACR criterion of the test metric.

ACR replicate measurements were made in the other two centers to investigate the intercenter reproducibility. Reproducibility was quantified by the standard deviation over the measurements in all three centers, $\sigma_{3c}$, and interpreted relative to the ACR tolerance using an intercenter coefficient of variation (iCV):

$$iCV = 100\% \frac{\sigma_{3c}}{c_{ACR}}. \tag{3}$$

**QIBA diffusion tests**

DWI measurements with both the prototype and standard setup were performed in one center (center 2) using the QIBA calibration phantom. The aim was to investigate the quality of DWI in a neck position (6 cm) using the prototype setup in comparison to the standard setup; and to investigate whether the prototype setup passes the QIBA acceptance criteria (qibawiki.rsna.org). The ice water cooled phantom contained vials of different media to allow the measurement of known apparent diffusion coefficients (ADC). A set of four repeat scans was performed with each setup using the QIBA benchmark protocol (single shot - echo planar imaging (SS-EPI); four b-values, 0, 500, 900, 2000 s/mm$^2$; diffusion mode, 3-scan trace; acquisition, transverse; signal averages, 1; field of view (FOV), 216x216x125 mm$^3$; slice thickness, 4 mm; spacing between slices, 5mm; repetition time (TR), 1000 ms; echo time (TE), 102 ms; pixel spacing, 1.125\1.125 mm). SNR in the DWI was analyzed based on the four repeat scans; an average signal image and a temporal noise image (temporal standard deviation across the four repeats) were calculated. SNR was derived as:

$$SNR = \frac{\bar{S}_{avg-ima}}{\bar{N}_{noise-ima}}, \tag{4}$$

where $\bar{S}_{avg-ima}$, an ROI-based mean value in the average signal image, and $\bar{N}_{noise-ima}$, an ROI-based mean value in the temporal noise image.

ADC bias (measured minus true ADC), percentage bias (ADC bias/true ADC), precision (spatial coefficient of variation, CV), short-term repeatability (temporal within-object coefficient of variation, wCV), and b-value dependence for different b-value combinations were analyzed for the prototype setup. The ADC b-value dependence was calculated as:

$$b\text{-}value\ dependence = \left| \frac{ADC_{b_0\text{-}b_{max1}} - ADC_{b_0\text{-}b_{max2}}}{ADC_{b_0\text{-}b_{max1}}} \right| 100\% , \tag{5}$$

where $ADC_{b_0\text{-}b_{max1}}$ and $ADC_{b_0\text{-}b_{max2}}$ are ADC values calculated from b-value combinations of $b_0$ (0 s/mm$^2$) and different $b_{max}$ ( $b_{max1}$, $b_{max2}$ equal 500, 900, or 2000 s/mm$^2$; while $b_{max1} \neq b_{max2}$). All calculations were ROI-based and according to QIBA guidelines.

Measurements and analysis were repeated with a clinical protocol for low distortion DWI (readout segmented EPI (RESOLVE); four b-values, 0, 50, 100, 800 s/mm$^2$; diffusion mode, 4-scan trace; acquisition, transverse; averages, 1; FOV, 216x216x125 mm$^3$; slice thickness, 4 mm; spacing between slices, 5 mm; TR, 4710 ms; TE, 73 ms; pixel spacing, 1.125/1.125 mm).

All image analysis was performed in Matlab r2020a (The MathWorks Inc., Natick, MA, USA).

**Volunteer test**

Two healthy volunteers underwent MR imaging after informed consent and in agreement with institutional ethics guidelines. Volunteers were imaged with both the prototype and the standard setup. From the clinical protocol of the EMINENCE study a 3D T2-SPACE (TR, 1300 ms; TE, 90 ms; slice thickness, 1 mm; number of averages, 1; flip angle, 110 deg; pixel spacing, 0.5\0.5 mm; FOV, 320x320x176 mm$^3$) and a T1-VIBE Dixon sequence for synthetic CT (TR, 4.05 ms; TE, 2.57 ms; slice thickness, 2 mm; number of averages, 2; flip angle, 9 deg; pixel spacing, 1.18\1.18 mm; FOV, 490x490x288 mm$^3$) were tested. SNR was compared between imaging setups for an ROI defined in normal muscle tissue in the anterior neck (6 cm distance from isocenter). SNR was calculated as the ratio of ROI mean signal over the ROI standard deviation.

## 3. Results

**ACR image quality tests**

ACR test results from the comparison of MR image quality with the prototype setup vs. standard setup are presented in Figure 2. The figure shows the results for each image quality test metric as a function of phantom position in SI direction. ACR acceptance criteria for spatial resolution AP (a), slice thickness accuracy (b), slice position accuracy (c, d), ghosting (g), and geometric accuracy AP (h) were met with both the prototype and standard setup at all scan positions. Criteria for spatial resolution and geometric accuracy in other directions were also met and are presented in Supplementary Figure 2. Criteria for PIU (e) and low-contrast object detectability scores (f) were met at all scan positions with the prototype setup, but not with the standard setup which failed at position 20 cm (thorax) for PIU and at all positions > 4 cm SI (neck/thorax) for detectability. Compared to the standard setup, MR signal with the prototype setup was on average increased by a factor of 1.03 at head (-12 – 0 cm), and 2.0 at neck/thorax bed positions (0 – 20 cm; Figure 2i). Similarly, SNR was increased by a factor of 1.03 at head, and 1.6 at neck/thorax positions (Figure 2k). The signal and SNR increase was statistically significant at neck/thorax (p=0.031 for both), but not at head positions.

**Repeatability and reproducibility**

Table 1 presents the monthly ACR test results of MR image quality with the prototype setup obtained in one center. Results for T1w and T2w MR image quality and geometric accuracy are listed per month. In addition, ACR acceptance criteria, monthly mean, standard deviation and the within-center coefficients of variation are shown. The prototype setup passed all ACR criteria for T1w MR, T2w MR and geometric accuracy in all tests, except for the PIU criterion of T1w and T2w MR which was not met in three of six tests. Thus, in total, 20 of 22 ACR criteria were met in all tests. Repeatability standard deviations were ≤ 1 mm for all differences in reference to distance, ≤ 9 percentage points for percentage differences and 0 spokes for low-contrast detectability scores. The monthly repeatability variation relative to the ACR criterion, wCV, was 4.9% on average.

Table 2 lists the ACR test results from all centers. Per center, all ACR acceptance criteria were met for all criteria except for PIU of T1w MR in center 2, where a small difference to the acceptance limit of 1% was found. Across centers, the upper limits of reproducibility standard deviations were similar to the ones for repeatability, within .01 mm, 1 percentage point and 1 spoke, except for slice position and a horizontal measure of geometric accuracy. These were up to .7 and 1.0 mm larger, respectively. The intercenter reproducibility variation relative to the ACR criterion, iCV, was 7.8% on average.

**QIBA diffusion tests**

SNR results from DWI phantom measurements with the prototype vs. standard setup are presented in Supplementary Figure 3, both for the QIBA benchmark sequence and the clinical RESOLVE sequence. The prototype setup passed the QIBA SNR criterion (SNR > 50) for all DW images of the clinical sequence and the benchmark sequence, except for the b2000 image. The standard setup passed the criterion for all DW images with the clinical sequence but failed the criterion for all images with the benchmark sequence. The SNR in the DW images was on average 1.7 (range: 1.66 – 1.87) times higher with the prototype setup compared to the standard setup.

Analysis of the ADC maps measured with the prototype setup yielded the results listed in Table 3. Benchmark sequence ADC maps with the prototype setup passed all QIBA acceptance criteria for ADC bias, ADC repeatability, and ADC b-value dependencies, but not for ADC precision. Similar results were observed for the clinical sequence, with the setup complying with all criteria but precision.

**Volunteer test**

T2-SPACE and T1-VIBE Dixon in-phase images of one volunteer are presented in Figure 3. Images acquired with prototype and standard setup are shown side-by-side. For both volunteers, the SNR in neck was 1.5 (T2-SPACE) and 1.7 (T1-VIBE Dixon) times higher for the prototype setup than for the standard setup.

## 4. Discussion

In this study we presented a novel prototype setup for PET/MR imaging in patients with head and neck cancer scheduled for radiotherapy. From a systematic analysis using the ACR phantom, our main finding was that the prototype setup increased the SNR at neck/thorax positions significantly compared to the standard setup by a factor of 1.6 (Fig. 2k, T1w MR, p=0.031). Imaging of healthy volunteers with sequences from a clinical protocol showed a similar increase in SNR in the neck by a factor of 1.5 to 1.7. The setup repeatedly met all ACR acceptance criteria of MR image quality in phantoms except for the image uniformity measure PIU. Interestingly, the setup complied with the criterion for low-contrast object detectability ($\geq$ 37 visible spokes out of 40 total) at all tested neck and thorax positions, whereas the standard setup did not (Fig. 2f). This finding may be attributed to the increase in SNR. Low-contrast detectability has earlier been shown to be a critical criterion for a pelvic radiotherapy setup presented by Wyatt et al. for a 3T PET/MR system (Signa PET/MR, GE Healthcare, Milwaukee, USA). Their study found scores of $22 \pm 2$ (T1w, values read from fig. 3 in (30)) or $15 \pm 1$ spokes (T2w) and suggested to increase number of signal averages during acquisition or to use noise reduction reconstruction techniques to improve SNR and thereby, detectability (30). The ACR compliant detectability scores with our prototype setup indicate higher image quality, superior compared to a standard setup in lower neck and upper thorax. This may benefit contouring accuracy of radiotherapy target and risk volumes or quantitative image feature calculations in such regions. Additional actions to increase SNR in clinical protocols as mentioned in (30) may still be considered an option, but not a necessity.

**Repeatability and reproducibility**

Another main finding was that MR image quality with the prototype setup was stable over three months and complied with ACR criteria in three different centers. For the ACR test metrics, the average reproducibility variation (7.8%) across centers was slightly larger than the average repeatability variation (4.9%) within a single center. This seems plausible as different imaging systems may add variation (33). Notably, geometric accuracy findings complied with the 2 mm acceptable distortion limit for MR-based radiotherapy (34). This points towards a robust performance of the proposed setup for multicenter imaging-based radiotherapy trials.

We could not find any literature where MR image quality with a radiotherapy setup for hybrid PET/MR has been investigated in monthly repeat scans or in a multicenter study before. However, Wyatt et al. studied monthly repeatability of MR image quality in a single center study for an anterior array coil and spine coil like in a pelvic setup but without using a flat tabletop or coil bridge. The authors documented monthly standard variations of up to 1.1 mm, .8 percentage points and 2 spokes maximum for distance, percentage and detectability scores (35). Our estimates of maximum variation seem in good agreement with (35) for distance and detectability measures, within .1 mm and 2 spokes, but less for image uniformity measure PIU. The higher PIU variation with our setup may be explained by the more complex coil configuration of three flexible coils and spine coil with varying distances to the phantom. Notably, at the head/neck position the spine coil covers only part of the posterior phantom side. In a pelvic setup the coil-phantom coverage and distance are more uniform. Also, in contrast to our study, flat tabletop and coil bridges were not present in the tests by Wyatt and colleagues resulting in a closer distance between phantom and coils, a higher coil fill factor, and likely, improved but optimistic estimates of SNR and image quality. We would recommend performing the tests with fully mounted radiotherapy setup including tabletop and coil holders or coil bridges for more realistic estimates of image quality.

**Selection of image quality metrics**

Some ACR metrics depend on the system rather than the coil setup, such as geometric accuracy and slice position, which are influenced by the system's B0 field inhomogeneity and/or gradient nonlinearity (36). Coil setup related image quality metrics include image uniformity, SNR and low-contrast object detectability. However, system related metrics were included in the analysis as these remain relevant for overall image quality performance as well as for radiotherapy planning purposes such as target and organ contouring or dose planning, which requires high spatial integrity of images.

**Different radiotherapy setups on the market**

Besides the pelvic setup by Wyatt et al., several setup designs with coil holders and body matrix coils have been proposed for PET/MR in radiotherapy patients (21,27–29,31). Besides reduced low-contrast object detectability, SNR in MR images has proven to be a major critical factor. Compared to diagnostic imaging, reductions in SNR of 25% up to 46% have been reported (27–30). These findings motivate improvements in setup design for better SNR and detectability, especially in neck and upper thorax. Two previous imaging setups have been proposed for patients with head and neck cancer (21,27,28,31), but neither setup has been optimized for neck or thorax imaging. In the studies conducted by Olin et al., Paulus et al., and Winter et al. the head and neck setup (Qfix, Avondale, PA, USA) consisted of a pair of head coil holders covering brain, head and chin but not the anterior neck/thorax (21,27,28), which is suboptimal for SNR and image quality in this region. An additional flex coil can be placed on neck and thorax to compensate for a lower SNR and image quality, but without a dedicated coil holder solution for a reproducible position it is not possible to correct for PET signal attenuation (21). Another head setup (Medibord) presented by Taeubert et al. did not include a neck/thorax coil setup or mask fixation devices (31). Paulus et al. proposed a whole-body setup, but it was not shown whether it would be compatible with neck and shoulder mask fixation (32). Brynolfsen et al. presented a radiotherapy coil setup for a GE PET/MR scanner, but imaging in head/neck/thorax as well as compatibility with head, neck and shoulder mask fixation was not investigated (29). A dedicated neck/thorax coil component as proposed with our study showed to improve MR signal, SNR and image quality in the neck/thorax region. This could benefit target contouring on anatomical MR images, the integration of functional MR techniques into radiotherapy, as well as developments in quantitative imaging biomarkers such as diffusion information derived from DWI, where signal and SNR are known to become critically low in images with high diffusion-weightings (37).

**Diffusion-weighted imaging**

With the diffusion phantom experiments in the neck bed position, we found that the prototype setup complied with the QIBA acceptance criterion for SNR in the DWI benchmark sequence for all b-values ≤ 900 s/mm$^2$. These results stand in contrast to the standard setup which did not fulfill the QIBA SNR criterion in any of the EPI b-value images. SNR was higher and complied with the QIBA criterion for both setups when using the RESOLVE sequence from the clinical protocol of the EMINENCE trial. Analysis of the ADC maps revealed that the prototype setup passed all ADC QIBA criteria except precision. Increasing the number of signal averages in the acquisition may be a simple method to further improve SNR and thus, ADC precision, but would come at the expense of longer scan time.

To our knowledge, this is the first study with a radiotherapy setup for PET/MR which follows the QIBA guidelines and compares diffusion phantom measures to QIBA benchmark criteria. QIBA tests were chosen as benchmark tests in this study, as these are a well-established consensus standard used to benchmark DWI performance of other novel hybrid MR systems such as the MR-Linac (33,38,39). The positive QIBA findings for SNR and ADC quality with the proposed setup increase the confidence that DWI information can be clinically used for radiotherapy contouring, integrated into treatment planning and used for developments of quantitative DWI biomarkers (37). The SNR in the b2000 image was below the criterion. However, to date, such high b-values are usually not used in head/neck cancer, and may be seen as less of a limitation (37).

**Limitations**

A main limitation in our study was that monthly ACR repeat measurements were only performed in one center, whereas results obtained in the two other centers relied on one-time measurements. Longitudinal repeats in all centers would enhance accuracy and robustness of the repeatability and reproducibility estimates. Moreover, longitudinal repeats were performed for three months only. Repeats over a 6- or 12-months period might give more insights into long term performance trends. However, in an ACR study with an experimental pelvic setup no monthly trends appeared over 12 months (35). Finally, QIBA experiments were only conducted in one center and repeatability was limited to short-time estimates. Monthly repeat and multicenter replicate tests of DWI would give more insights into long term repeatability and reproducibility. Performance assessment of PET was beyond the scope of this work. Feasibility of PET attenuation correction has been demonstrated for radiotherapy setups elsewhere (27,31), and will be evaluated for our setup in future work.

**Conclusions**

A novel prototype radiotherapy setup for hybrid PET/MR imaging in patients with head and neck cancer was implemented in three centers. Phantom experiments suggest higher SNR and better detectability in neck and upper thorax compared to a standard radiotherapy setup. Proof-of-principle was demonstrated in healthy volunteers. Compliance with nearly all ACR and QIBA acceptance criteria as well as good repeatability and reproducibility indicate robust usability of the setup for multicenter trials. The proposed setup may thus advance the usage of PET/MR in radiotherapy and benefit developments of quantitative imaging biomarkers for personalized treatment of head and neck cancer.

# References

1. Gurney-Champion OJ, Mahmood F, van Schie M, Julian R, George B, Philippens MEP, et al. Quantitative imaging for radiotherapy purposes. Radiother Oncol. 2020;146:66–75.

2. Grégoire V, Guckenberger M, Haustermans K, Lagendijk JJW, Ménard C, Pötter R, et al. Image guidance in radiation therapy for better cure of cancer. Mol Oncol. 2020;14(7):1470–91.

3. Chandarana H, Wang H, Tijssen R h. n., Das IJ. Emerging role of MRI in radiation therapy. J Magn Reson Imaging. 2018;48(6):1468–78.

4. Alterio D, Marvaso G, Ferrari A, Volpe S, Orecchia R, Jereczek-Fossa BA. Modern radiotherapy for head and neck cancer. Semin Oncol. 2019; 46(3):233–245.

5. Min M, Lin P, Lee MT, Shon IH, Lin M, Forstner D, et al. Prognostic role of metabolic parameters of 18F-FDG PET-CT scan performed during radiation therapy in locally advanced head and neck squamous cell carcinoma. Eur J Nucl Med Mol Imaging. 2015;42(13):1984–94.

6. Thorwarth D, Welz S, Mönnich D, Pfannenberg C, Nikolaou K, Reimold M, et al. Prospective evaluation of a tumor control probability model based on dynamic 18F-FMISO PET for head-and-neck cancer radiotherapy. J Nucl Med. 2019;60(12):1698–1704.

7. Driessen JP, Caldas-Magalhaes J, Janssen LM, Pameijer FA, Kooij N, Terhaard CHJ, et al. Diffusion-weighted MR imaging in laryngeal and hypopharyngeal carcinoma: association between apparent diffusion coefficient and histologic findings. Radiology. 2014;272(2):456–63.

8. Jensen K, Al-Farra G, Dejanovic D, Eriksen JG, Loft A, Hansen CR, et al. Imaging for Target Delineation in Head and Neck Cancer Radiotherapy. Semin Nucl Med. 2021;51(1):59–67.

9. Leibfarth S, Eckert F, Welz S, Siegel C, Schmidt H, Schwenzer N, et al. Automatic delineation of tumor volumes by co-segmentation of combined PET/MR data. Phys Med Biol. 2015;60(14):5399–412.

10. Cao Y, Aryal M, Li P, Lee C, Schipper M, Hawkins PG, et al. Predictive Values of MRI and PET Derived Quantitative Parameters for Patterns of Failure in Both p16+ and p16-High Risk Head and Neck Cancer. Front Oncol. 2019;9:1118.

11. Mes SW, van Velden FHP, Peltenburg B, Peeters CFW, te Beest DE, van de Wiel MA, et al. Outcome prediction of head and neck squamous cell carcinoma by MRI radiomic signatures. Eur Radiol. 2020;30(11):6311–21.

12. Crispin-Ortuzar M, Apte A, Grkovski M, Oh JH, Lee NY, Schöder H, et al. Predicting hypoxia status using a combination of contrast-enhanced computed tomography and [18F]-Fluorodeoxyglucose positron emission tomography radiomics features. Radiother Oncol. 2018;127(1):36–42.

13. Welz S, Paulsen F, Pfannenberg C, Reimold M, Reischl G, Nikolaou K, et al. Dose escalation to hypoxic subvolumes in head and neck cancer: A randomized phase II study using dynamic [18F]FMISO PET/CT. Radiother Oncol. 2022;171:30–6.

14. Yu T ting, Lam S kit, To L hang, Tse K yan, Cheng N yi, Fan Y nam, et al. Pretreatment Prediction of Adaptive Radiation Therapy Eligibility Using MRI-Based Radiomics for Advanced Nasopharyngeal Carcinoma Patients. Front Oncol. 2019;9:1118.

15. Thorwarth D, Leibfarth S, Mönnich D. Potential role of PET/MRI in radiotherapy treatment planning. Clin Transl Imaging. 2013 Feb;1(1):45–51.

16. Zhu T, Das S, Wong TZ. Integration of PET/MR Hybrid Imaging into Radiation Therapy Treatment. Magn Reson Imaging Clin N Am. 2017;25(2):377–430.

17. Bailey DL, Pichler BJ, Gückel B, Antoch G, Barthel H, Bhujwalla ZM, et al. Combined PET/MRI: Global Warming—Summary Report of the 6th International Workshop on PET/MRI, March 27–29, 2017, Tübingen, Germany. Mol Imaging Biol. 2018;20(1):4–20.

18. Yan Q, Yan X, Yang X, Li S, Song J. The use of PET/MRI in radiotherapy. Insights Imaging. 2024;15:63.

19. Wang Y, Liu C, Zhang X, Deng W. Synthetic CT Generation Based on T2 Weighted MRI of Nasopharyngeal Carcinoma (NPC) Using a Deep Convolutional Neural Network (DCNN). Front Oncol. 2019;9:1333

20. Dinkla AM, Florkow MC, Maspero M, Savenije MHF, Zijlstra F, Doornaert PAH, et al. Dosimetric evaluation of synthetic CT for head and neck radiotherapy generated by a patch-based three-dimensional convolutional neural network. Med Phys. 2019;46(9):4095–104.

21. Olin AB, Hansen AE, Rasmussen JH, Ladefoged CN, Berthelsen AK, Håkansson K, et al. Feasibility of Multiparametric Positron Emission Tomography/Magnetic Resonance Imaging as a One-Stop Shop for Radiation Therapy Planning for Patients with Head and Neck Cancer. Int J Radiat Oncol. 2020;108(5):1329–38.

22. Ahangari S, Hansen NL, Olin AB, Nøttrup TJ, Ryssel H, Berthelsen AK, et al. Toward PET/MRI as one-stop shop for radiotherapy planning in cervical cancer patients. Acta Oncol. 2021;60(8):1045–53.

23. Fortunati V, Verhaart RF, Verduijn GM, van der Lugt A, Angeloni F, Niessen WJ, et al. MRI integration into treatment planning of head and neck tumors: Can patient immobilization be avoided? Radiother Oncol. 2015;115(2):191–4.

24. Paulus DH, Quick HH. Hybrid Positron Emission Tomography/Magnetic Resonance Imaging: Challenges, Methods, and State of the Art of Hardware Component Attenuation Correction. Invest Radiol. 2016;51(10):624–34.

25. Schmidt MA, Payne GS. Radiotherapy planning using MRI. Phys Med Biol. 2015;60(22):R323–61.

26. Gruber B, Froeling M, Leiner T, Klomp DWJ. RF coils: A practical guide for nonphysicists. J Magn Reson Imaging. 2018;48(3):590–604.

27. Paulus DH, Thorwath D, Schmidt H, Quick HH. Towards integration of PET/MR hybrid imaging into radiation therapy treatment planning. Med Phys. 2014;41(7):072505.

28. Winter RM, Leibfarth S, Schmidt H, Zwirner K, Mönnich D, Welz S, et al. Assessment of image quality of a radiotherapy-specific hardware solution for PET/MRI in head and neck cancer patients. Radiother Oncol. 2018;128(3):485–91.

29. Brynolfsson P, Axelsson J, Holmberg A, Jonsson JH, Goldhaber D, Jian Y, et al. Technical Note: Adapting a GE SIGNA PET/MR scanner for radiotherapy. Med Phys. 2018;45(8):3546–50.

30. Wyatt JJ, Howell E, Lohezic M, McCallum HM, Maxwell RJ. Evaluating the image quality of combined positron emission tomography-magnetic resonance images acquired in the pelvic radiotherapy position. Phys Med Biol. 2021;66(3):035018.

31. Taeubert L, Berker Y, Beuthien-Baumann B, Hoffmann AL, Troost EGC, Kachelrieß M, et al. CT-based attenuation correction of whole-body radiotherapy treatment positioning devices in PET/MRI hybrid imaging. Phys Med Biol. 2020;65(23):23NT02.

32. Paulus DH, Oehmigen M, Grueneisen J, Umutlu L, Quick HH. Whole-body hybrid imaging concept for the integration of PET/MR into radiation therapy treatment planning. Phys Med Biol. 2016;61(9):3504–20.

33. van Houdt PJ, Kallehauge JF, Tanderup K, Nout R, Zaletelj M, Tadic T, et al. Phantom-based quality assurance for multicenter quantitative MRI in locally advanced cervical cancer. Radiother Oncol. 2020;153:114–21.

34. Weygand J, Fuller CD, Ibbott GS, Mohamed ASR, Ding Y, Yang J, et al. Spatial Precision in Magnetic Resonance Imaging–Guided Radiation Therapy: The Role of Geometric Distortion. Int J Radiat Oncol. 2016;95(4):1304–16.

35. Wyatt JJ, McCallum HM, Maxwell RJ. Developing quality assurance tests for simultaneous Positron Emission Tomography – Magnetic Resonance imaging for radiotherapy planning. Phys Imaging Radiat Oncol. 2022;22:28–35.

36. Adjeiwaah M, Garpebring A, Nyholm T. Sensitivity analysis of different quality assurance methods for magnetic resonance imaging in radiotherapy. Phys Imaging Radiat Oncol. 2020;13:21–7.

37. Leibfarth S, Winter RM, Lyng H, Zips D, Thorwarth D. Potentials and challenges of diffusion-weighted magnetic resonance imaging in radiotherapy. Clin Transl Radiat Oncol. 2018;13:29–37.

38. van Houdt PJ, Saeed H, Thorwarth D, Fuller CD, Hall WA, McDonald BA, et al. Integration of quantitative imaging biomarkers in clinical trials for MR-guided radiotherapy: Conceptual guidance for multicentre studies from the MR-Linac Consortium Imaging Biomarker Working Group. Eur J Cancer. 2021;153:64–71.

39. Kooreman ES, van Houdt PJ, Nowee ME, van Pelt VWJ, Tijssen RHN, Paulson ES, et al. Feasibility and accuracy of quantitative imaging on a 1.5 T MR-linear accelerator. Radiother Oncol. 2019;133:156–62.

# Tables

**Table 1:** Monthly repeatability of MR image quality with the prototype setup.

| Test | Metric | ACR criterion | Month 1 | Month 2 | Month 3 | Mean | $\sigma_{3m}$ | wCV (%) |
|---|---|---|---|---|---|---|---|---|
| T1w MR image quality | Spatial resolution AP (mm) | ≤ 1.0 | 1.0 | 0.9 | 0.9 | 0.93 | 0.06 | 6 |
| | Spatial resolution LR (mm) | ≤ 1.0 | 1.0 | 0.9 | 0.9 | 0.93 | 0.06 | 6 |
| | Slice thickness accuracy (mm) | 5.0 ± 1.0 | 5.7 | 5.0 | 5.4 | 5.4 | 0.4 | 8 |
| | Slice position accuracy slice 1 (mm) | ≤ 5.0 | 1.5 | 1.9 | 1.0 | 1.5 | 0.5 | 10 |
| | Slice position accuracy slice 11 (mm) | ≤ 5.0 | 3.9 | 2.9 | 4.9 | 3.9 | 1.0 | 20 |
| | Low-contrast detectability (#spokes) | ≥ 37 | 38 | 38 | 38 | 38 | 0 | 0 |
| | PIU (%) | ≥ 80 | **79** | 92 | 91 | 87 | 7 | 9 |
| | Ghosting ratio (%) | ≤ 3 | 0.1 | 0.2 | 0.3 | 0.2 | 0.1 | 3 |
| | SNR | - | - | 799 | 921 | 860 | 86 | - |
| T2w MR image quality | Spatial resolution AP (mm) | ≤ 1.0 | 0.9 | 0.9 | 0.9 | 0.9 | 0 | 0 |
| | Spatial resolution LR (mm) | ≤ 1.0 | 0.9 | 0.9 | 0.9 | 0.9 | 0 | 0 |
| | Slice thickness accuracy (mm) | 5.0 ± 1.0 | 4.6 | 4.9 | 5.4 | 5.0 | 0.4 | 9 |
| | Slice position accuracy slice 1 (mm) | ≤ 5.0 | 2.4 | 2.0 | 1.0 | 1.8 | 0.8 | 15 |
| | Slice position accuracy slice 11 (mm) | ≤ 5.0 | 2.9 | 2.9 | 3.9 | 3.3 | 0.6 | 11 |
| | Low-contrast detectability | ≥ 37 | 38 | 38 | 38 | 38 | 0 | 0 |
| | PIU (%) | ≥ 80 | 94 | **77** | **79** | 83 | 9 | 11 |
| Geometric accuracy | Length SI (mm) | 148 ± 3 | 147.0 | 146.5 | 146.5 | 146.6 | 0.3 | 0.2 |
| | Diameter AP in slice 1 (mm) | 190 ± 3 | 189.9 | 189.5 | 189.5 | 189.6 | 0.3 | 0.1 |
| | Diameter LR in slice 1 (mm) | 190 ± 3 | 190.9 | 192.4 | 192.4 | 191.9 | 0.8 | 0.4 |
| | Diameter AP in slice 5 (mm) | 190 ± 3 | 189.5 | 188.5 | 188.5 | 189.1 | 0.6 | 0.3 |
| | Diameter LR in slice 5 (mm) | 190 ± 3 | 189.9 | 190.4 | 190.4 | 190.3 | 0.3 | 0.1 |
| | Diameter LLUR in slice 5 (mm) | 190 ± 3 | 188.5 | 189.2 | 189.2 | 189.0 | 0.4 | 0.2 |
| | Diameter ULLR in slice 5 (mm) | 190 ± 3 | 189.2 | 190.6 | 189.2 | 189.7 | 0.8 | 0.4 |

Abbrev.: ACR = American College of Radiology; T1w = T1-weighted; CV = coefficient of variation; PIU = percent image uniformity; SNR = signal-to-noise ratio; $\sigma_{3m}$ = three-month standard deviation; SI = superior-inferior; AP = anterior-posterior; LR = left-right; LLUR = lower-left to upper-right diagonal; ULLR = upper-left to lower-right diagonal.

Results which did not comply with the ACR criterion are highlighted (bold font). Note that wCV is given in percent.

**Table 2:** Intercenter reproducibility of MR image quality with the prototype setup.

| Test | Metric | ACR criterion | Center 1 | Center 2 | Center 3 | Mean | $\sigma_{3c}$ | iCV (%) |
|------|--------|--------------|----------|----------|----------|------|---------|---------|
| T1w MR image quality | Spatial resolution AP (mm) | ≤ 1.0 | 0.9 | 1.0 | 1.0 | 0.97 | 0.06 | 6 |
| | Spatial resolution LR (mm) | ≤ 1.0 | 0.9 | 1.0 | 1.0 | 0.97 | 0.06 | 6 |
| | Slice thickness accuracy (mm) | 5.0 ± 1.0 | 4.9 | 5.7 | 5.0 | 5.2 | 0.5 | 9 |
| | Slice position accuracy slice 1 (mm) | ≤ 5.0 | 1.0 | 1.5 | 3.9 | 2.1 | 1.6 | 31 |
| | Slice position accuracy slice 11 (mm) | ≤ 5.0 | 2.9 | 3.9 | 2.0 | 2.9 | 1.0 | 20 |
| | Low-contrast detectability | ≥ 37 | 38 | 38 | 38 | 38 | 0 | 0 |
| | PIU (%) | ≥ 80 | 91 | **79** | 96 | 89 | 9 | 11 |
| | Ghosting ratio (%) | ≤ 3 | 0.3 | 0.1 | 0.2 | 0.2 | 0.1 | 3 |
| | SNR | - | 1045 | 799 | - | 922 | 174 | - |
| T2w MR image quality | Spatial resolution AP (mm) | ≤ 1.0 | 0.9 | 0.9 | 0.9 | 0.9 | 0 | 0 |
| | Spatial resolution LR (mm) | ≤ 1.0 | 0.9 | 0.9 | 0.9 | 0.9 | 0 | 0 |
| | Slice thickness accuracy (mm) | 5.0 ± 1.0 | 4.7 | 4.6 | 5.6 | 5.0 | 0.6 | 12 |
| | Slice position accuracy slice 1 (mm) | ≤ 5.0 | 0.5 | 2.4 | 3.9 | 2.3 | 1.7 | 34 |
| | Slice position accuracy slice 11 (mm) | ≤ 5.0 | 3.9 | 2.9 | 1.5 | 2.8 | 1.2 | 25 |
| | Low-contrast detectability | ≥ 37 | 38 | 38 | 37 | 38 | 0.6 | 2 |
| | PIU (%) | ≥ 80 | 80 | 94 | 95 | 90 | 8 | 10 |
| Geometric accuracy | Length SI (mm) | 148 ± 3 | 147.0 | 147.0 | 146.5 | 146.8 | 0.3 | 0.2 |
| | Diameter AP in slice 1 (mm) | 190 ± 3 | 191.4 | 189.9 | 190.4 | 190.6 | 0.8 | 0.4 |
| | Diameter LR in slice 1 (mm) | 190 ± 3 | 192.4 | 190.9 | 188.5 | 190.6 | 2.0 | 1.0 |
| | Diameter AP in slice 5 (mm) | 190 ± 3 | 190.4 | 189.5 | 190.4 | 190.1 | 0.6 | 0.3 |
| | Diameter LR in slice 5 (mm) | 190 ± 3 | 190.4 | 189.9 | 188.5 | 189.6 | 1.0 | 0.5 |
| | Diameter LLUR in slice 5 (mm) | 190 ± 3 | 189.2 | 188.5 | 188.5 | 188.8 | 0.4 | 0.2 |
| | Diameter ULLR in slice 5 (mm) | 190 ± 3 | 190.6 | 189.2 | 189.2 | 189.7 | 0.8 | 0.4 |

Abbrev.: ACR = American College of Radiology; T1w = T1-weighted; CV = coefficient of variation; PIU = percent image uniformity; SNR = signal-to-noise ratio; σ_3c = three-center standard deviation; SI = superior-inferior; AP = anterior-posterior; LR = left-right; LLUR = lower-left to upper-right diagonal; ULLR = upper-left to lower-right diagonal.

Results which did not comply with the ACR criterion are highlighted (bold font). Note that iCV is given in percent.

**Table 3:** DWI image quality with the prototype setup for two different sequences.

| Test | Metric | QIBA criterion | Benchmark sequence | Clinical sequence |
|---|---|---|---|---|
| DWI image quality | ADC bias (x10$^{-3}$ mm²/s) | ≤ 0.04 | -0.02 ± 0.01 | -0.02 ± 0.01 |
| | ADC bias (%) | ≤ 3.6 | -1.9 | -1.4 |
| | ADC repeatability (%) | ≤ 0.5 | 0.4 | 0.4 |
| | ADC precision (%) | ≤ 2 | **4.8** | **2.2** |
| | SNR in b0 image | ≥ 50 | 72 ± 4 | 113 ± 6 |
| | ADC b-value dependence b0-b500 vs. b0-b900 (%)* | < 2 | 1.0 | NA |
| | ADC b-value dependence b0-b900 vs. b0-b2000 (%)* | < 2 | 0.4 | NA |
| | ADC b-value dependence b0-b500 vs. b0-b2000 (%)* | < 2 | 1.4 | NA |

Abbrev.: DWI = diffusion-weighted MR imaging; ADC = apparent diffusion coefficient; SNR = signal-to-noise ratio; QIBA = Quantitative Imaging Biomarker Alliance; Benchmark sequence = EPI sequence; Clinical sequence = RESOLVE sequence.

*ADC b-value dependence was not evaluated for the clinical sequence, as different b-values were used.

Results which do not comply with the QIBA criteria are highlighted (bold font).

# Figure legends

## Figure 1

Different hardware components of the radiotherapy setups. a: Flat tabletop. b: Base plate for mask fixation and head/neck positioning aid. c: Head coil holders. d: Full head-only setup with two coils (standard setup). e: Novel coil holder for neck/thorax, together with the standard head coil holder and mask fixation devices. f: Full prototype setup with all three coils.

## Figure 2

MR image quality with the prototype setup (blue) compared to the standard setup (orange) at different positions along the table (SI direction). The ACR phantom and T1w MR sequence were used. ACR criteria are indicated as reference (black dotted lines). Error bars in (i, k) depict 95% confidence intervals, and are similar to or smaller than marker size. Abbrev.: AP = anterior-posterior; PIU = percent image uniformity; SNR = signal-to-noise ratio.

## Figure 3

T2-SPACE (a-d) and in-phase T1-VIBE Dixon images (e-f) in a healthy volunteer. Images acquired with prototype setup (a,c;e,g) and standard setup (b,d;f,h) are shown side-by-side. The transverse neck images (a,b;e,f) are located at 6 cm SI position. Images in each row are displayed with same window width and level.

# Figures

**Figure 1:**

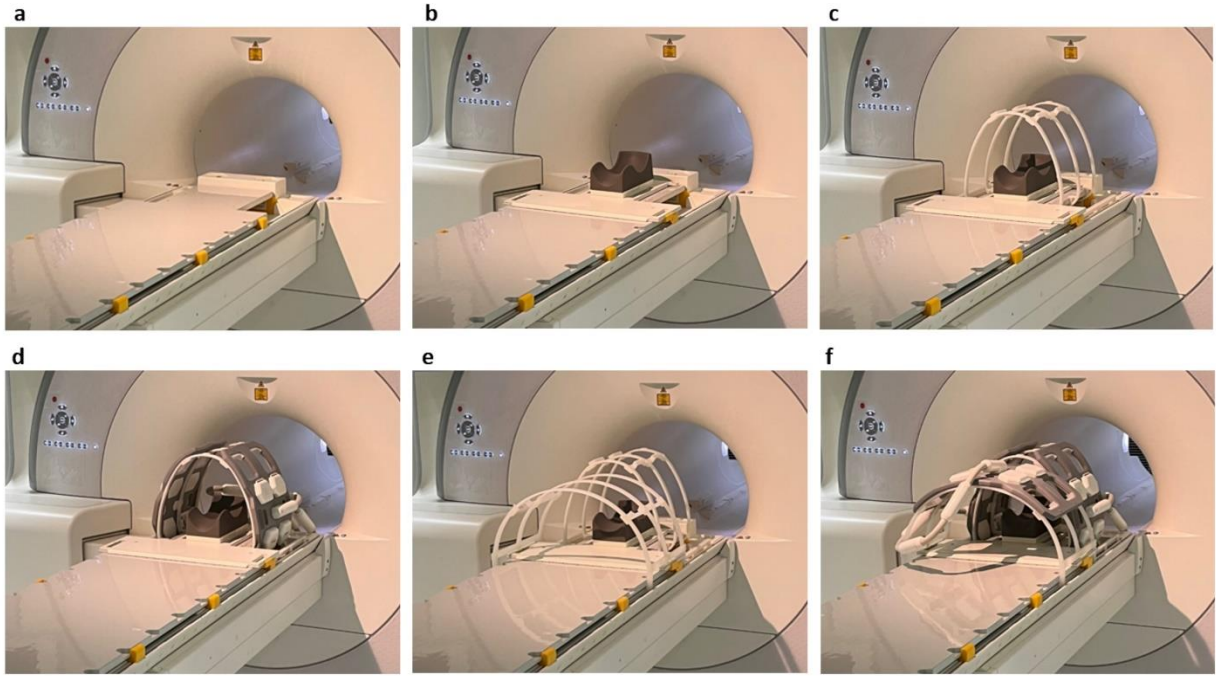

**Figure 2:**

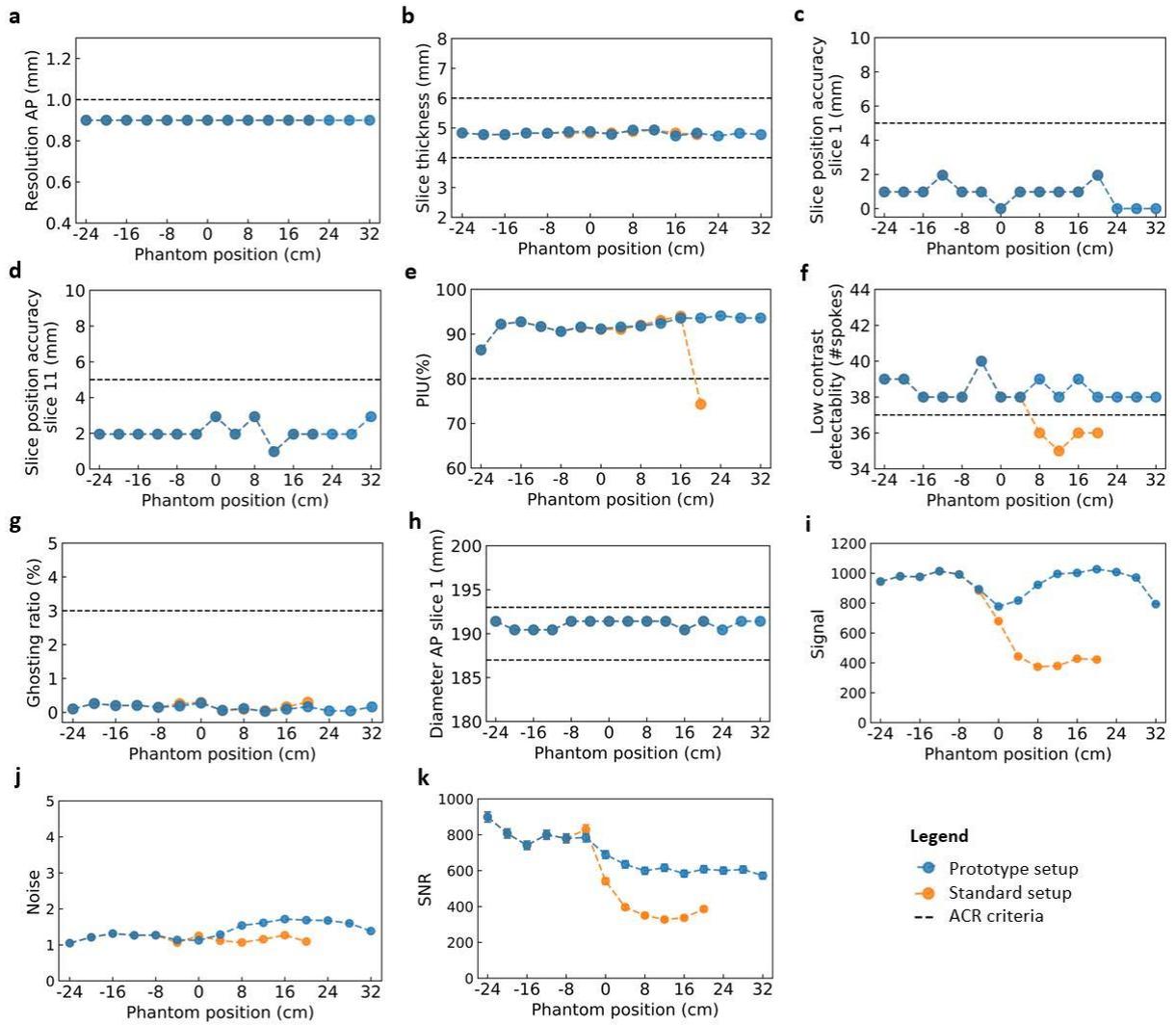

**Figure 3:**

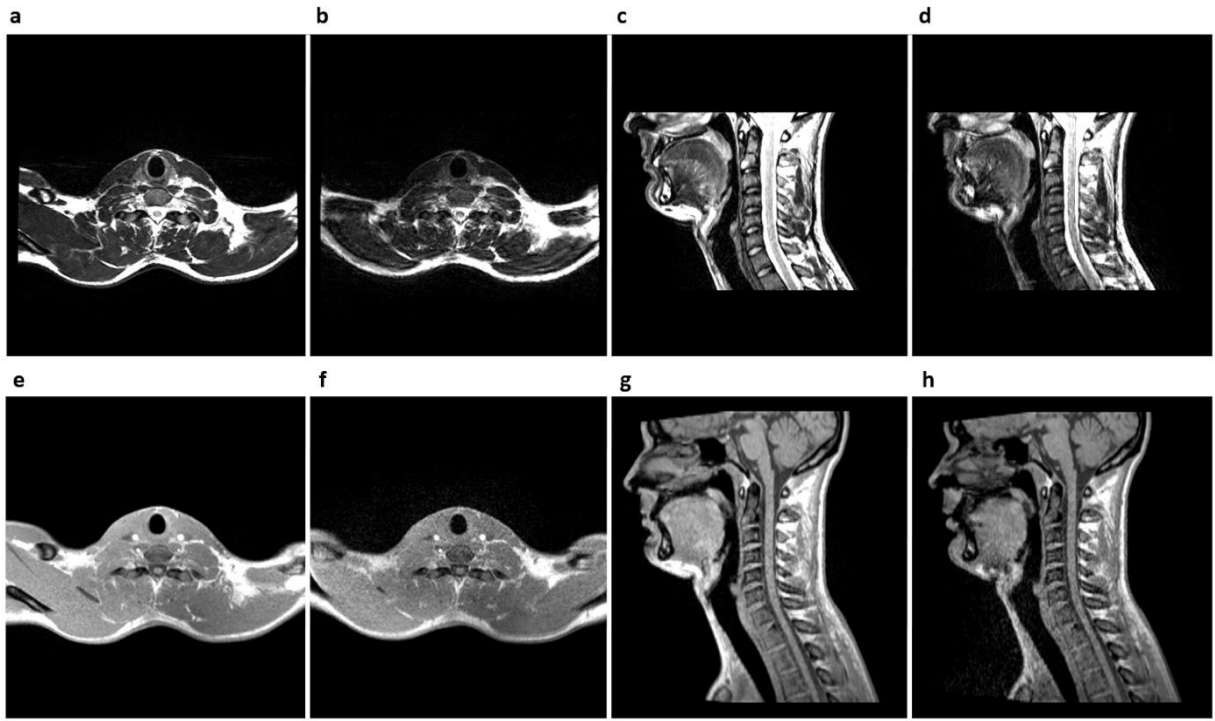

# Supplementary Material

## Supplementary Figures

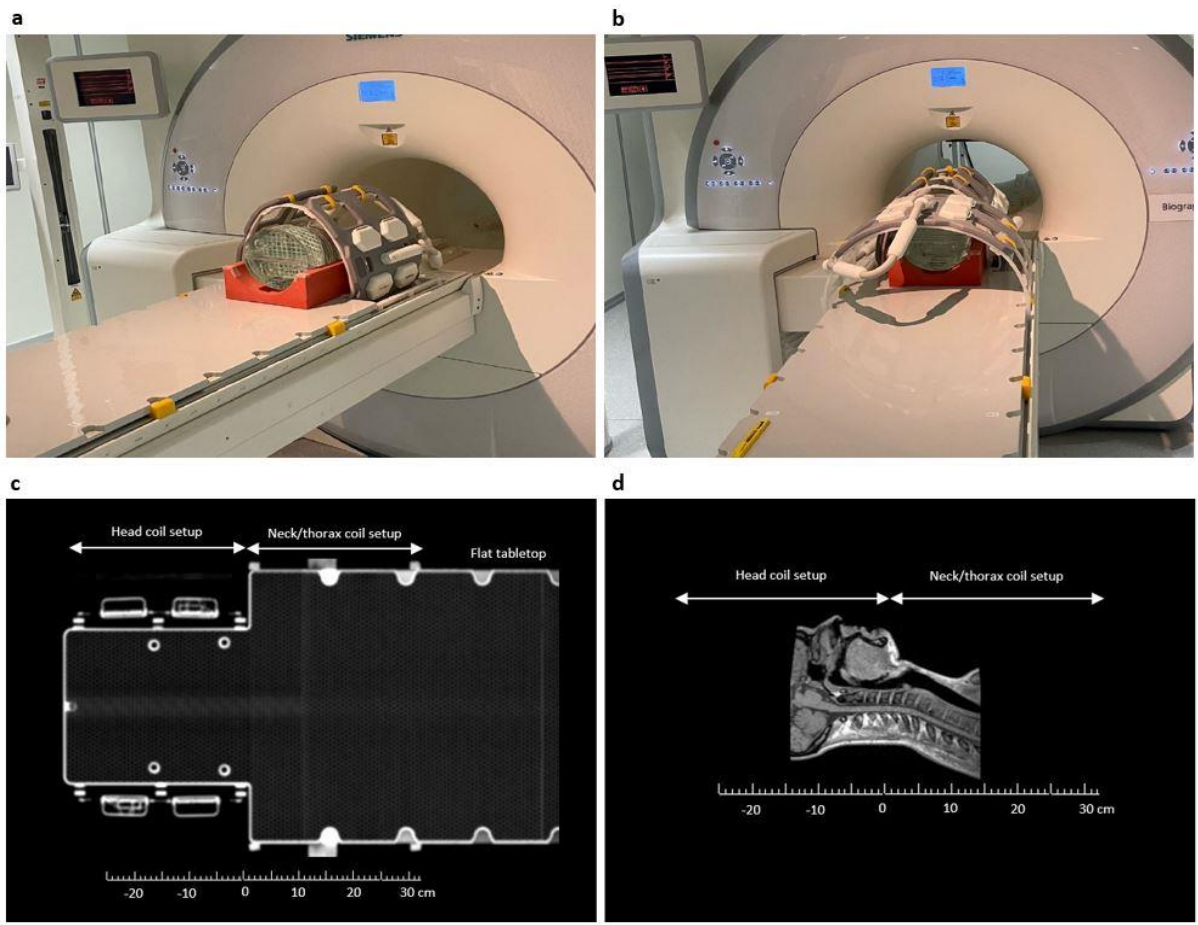

**Supplementary Figure 1:** Experimental setup with the ACR phantom placed in the standard setup (a) and the prototype setup (b). (c)-(d): schematic overview to illustrate phantom positioning with respect to the dimensions of the radiotherapy setups and the patient. The coronal image of the radiotherapy hardware in (c) was derived from a CT scan, the sagittal MR image in (d) was taken in a volunteer. Images are true to scale. Analysis was focused on phantom positions from -12 to 0, 0 to 12 and 12 to 20 cm SI which were considered as head, neck and upper thorax positions, respectively. This range is most relevant for head/neck cancer patients, while the prototype setup can in principle be used for imaging from brain to lower thorax.

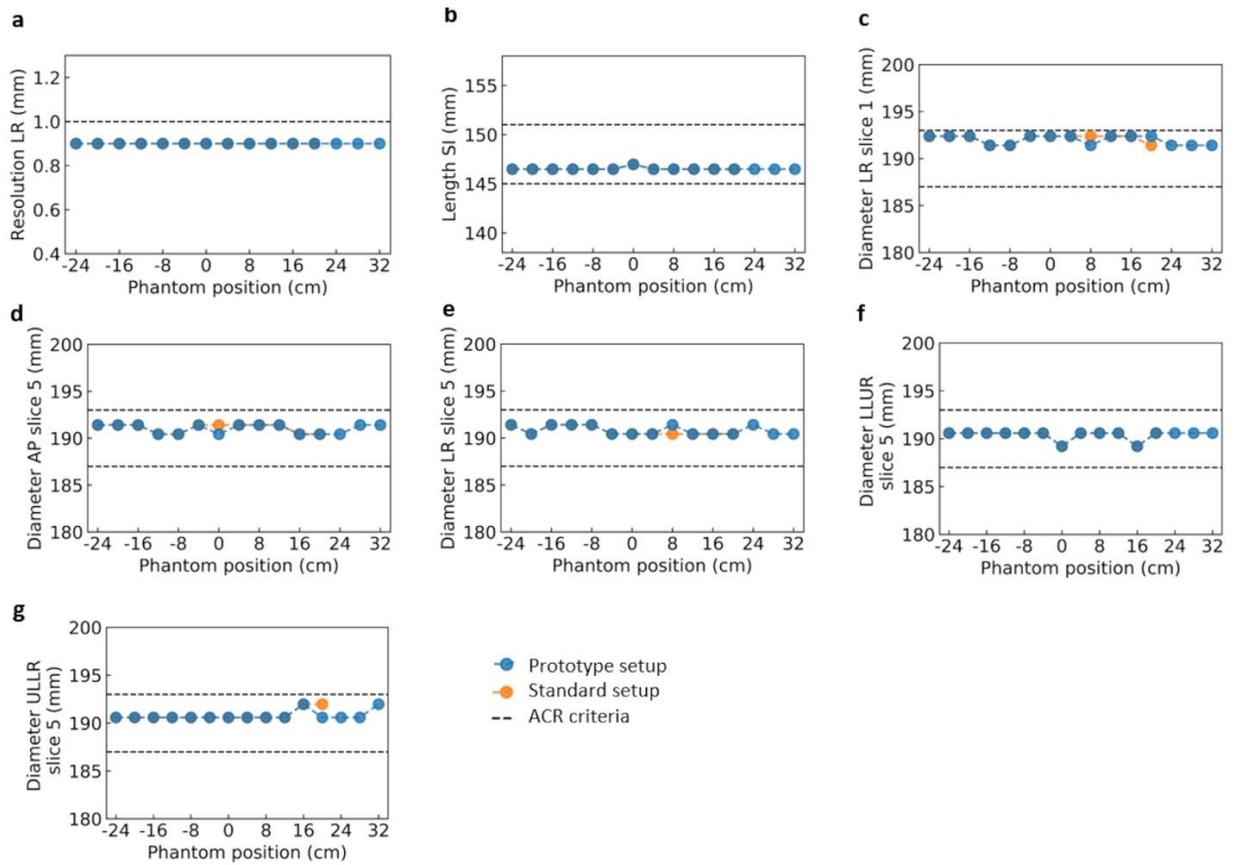

**Supplementary Figure 2:** Spatial resolution (a) and geometric accuracy in different directions (b-g) with head/neck/thorax prototype setup (blue) as compared to the standard head-only setup (orange) evaluated with the ACR phantom. ACR criteria are indicated by dotted black lines. Abbrev.: LR, left-right; SI, superior-inferior; AP, anteroposterior; sl., image slice; LLUR, lower left to upper right; ULLR, upper left to lower right.

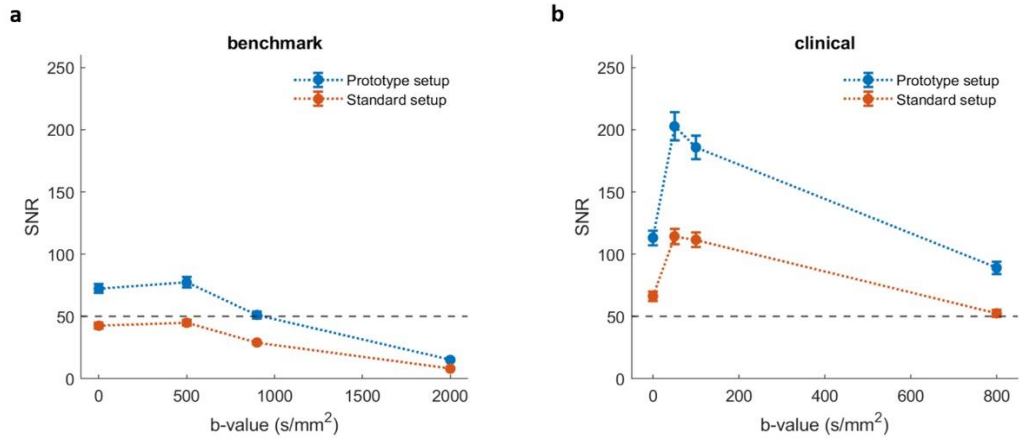

**Supplementary Figure 3:** Signal-to-noise ratio (SNR) in diffusion-weighted images with head/neck/thorax prototype (blue) and standard head-only setup (orange), plotted as function of b-value. a: SNR with the benchmark EPI sequence, b: with the clinical RESOLVE sequence. Data was measured with the QIBA diffusion phantom at a neck position. QIBA acceptance criterion for minimum SNR is indicated as dotted line. SNR of b0 appears reduced compared to b-values > 0, as the 3-scan/4-scan trace method enhances the SNR for b > 0.